# Photoluminescence of tetrahedral quantum-dot quantum wells

V. A. Fonoberov[a,b], E. P. Pokatilov[a], V. M. Fomin[c,d,a] and J. T. Devreese[c,d]

[a]*Laboratory of Physics of Multilayer Structures, Department of Theoretical Physics, State University of Moldova, MD-2009 Kishinev, Moldova*
[b]*Nano-Device Laboratory, Department of Electrical Engineering, University of California – Riverside, Riverside, California 92521-0425, USA*
[c]*Theoretische Fysica van de Vaste Stoffen (TFVS), Departement Natuurkunde, Universiteit Antwerpen, B-2610 Antwerpen, Belgium*
[d]*Technische Universiteit Eindhoven, NL-5600 MB Eindhoven, The Netherlands*

Taking into account the tetrahedral shape of a quantum dot quantum well (QDQW) when describing excitonic states, phonon modes and the exciton-phonon interaction in the structure, we obtain within a non-adiabatic approach a quantitative interpretation of the photoluminescence (PL) spectrum of a single CdS/HgS/CdS QDQW. We find that the exciton ground state in a tetrahedral QDQW is bright, in contrast to the dark ground state for a spherical QDQW.

## 1. Introduction

Preparation of the CdS quantum dots with HgS quantum wells (QDQW's) was described in Ref. [1]. Such structures possess greater photochemical stability as compared to ordinary quantum dots (QD's). Excitons, localized in the HgS quantum well, are separated from the localized surface levels by the CdS shell and, as result, have longer lifetime and improved quantum yield/photoresponse. It has been revealed that QDQW's are preferentially tetrahedral particles [2] with zincblende crystal lattice. Using confocal optical microscopy, photoluminescence (PL) from a single QDQW has been measured [3]. Multiband **k·p** theory for spherical QDQW's gives an s-type electron and a p-type hole ground states leading to the dark exciton ground state. In Ref. [4], we have reported the first quantitative interpretation of the PL spectrum of a QDQW.

## 2. Electron, hole and exciton states in QDQW

The specific features of the QDQW are (i) considerable difference of the effective-mass parameters between the well and the barrier materials on the interfaces of the thin HgS shell and (ii) coupling of the conduction band with the valence band in HgS due to a small value of the band gap. In order to take the above features into account, we use the non-symmetrized 8-band Hamiltonian, which was derived earlier by us for quantum-dot heterostructures [5]. To optimize the numerical calculations within the framework of the method described in Ref. [5] we have derived the 2-band electron Hamiltonian $H_2 = H_e$ and the 6-band hole Hamiltonian $H_6 = H_h$ from the general 8-band Hamiltonian.

We would like to emphasize that the influence of the valence band on the conduction band is included in the parameters of the electron Hamiltonian, and the influence of the conduction band on the valence band is included in the parameters of the hole Hamiltonian. The effective-mass parameters of HgS entering the Hamiltonians $H_e$ and $H_h$ are calculated as a function of the energy of the hole and the electron ground states, correspondingly. The HgS shell forms a potential well for both the electron and the hole.

Dielectric mismatch between the QDQW shells leads to the appearance of the energy of self-interaction for the electron and the hole $V_{s-a}$ and considerably changes the potential of the electron-hole interaction $V_{int}$. Taking into account the high quantization energy of the electron states, we consider the exciton wave function as a product of the wave function of the electron ground state $\varphi(\vec{r}_e)$ and the wave function of the hole, found as a solution of the Schrödinger equation with the Hamiltonian, averaged over the electron wave function:

$$<\varphi|H_{exc}|\varphi> = E_e - [H_h - V_{s-a}(\vec{r}_h)] + <\varphi|V_{int}(\vec{r}_e,\vec{r}_h)|\varphi> , \quad (1)$$

where $E_e$ is the electron ground state energy including the energy of the self-interaction.

The finite-difference method has been used to calculate the electron and hole states from the Hamiltonians $H_e$ and $H_h$ as well as the exciton energies from the Hamiltonian (1). For the ground states of the electron, the hole and the exciton the following values of the energy have been obtained: $E_{e0} = 1329$ meV, $E_{h0} = -591$ meV, $E_{exc0} = 1857$ meV. The calculated energy of the exciton ground state is in a good agreement with the experimental data [2,3]. The Stokes shift, obtained from the difference between the energies of the first excited and ground exciton states, is equal to 18 meV. It practically coincides with the experimentally observed value of 19 meV [2]. Figure 1 shows the intensity of the exciton levels. A remarkable effect of the QDQW's shape is that the exciton in the ground state is bright for the tetrahedral QDQW, while it is dark for the spherical QDQW.

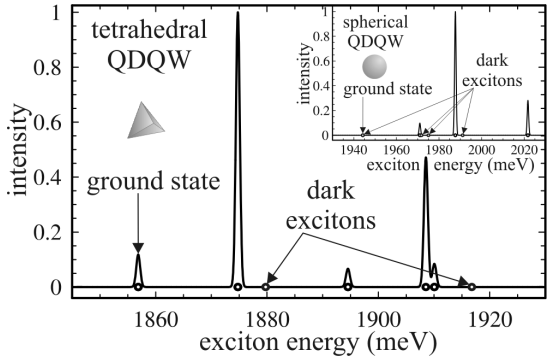

Fig.1. Normalized intensities of the lowest exciton levels in the optical absorption spectrum of a tetrahedral QDQW (the insert gives the intensities for a spherical QDQW). Spectral lines are broadened by a Gaussian to enhance visualization.

## 3. Electron-phonon interaction in the CdS/HgS/CdS QDQW

Expanding the lattice polarization $W(\vec{r},t)$ in terms of the orthonormalized modes $\vec{w}_\lambda(\vec{r})$:

$$\vec{W}(\vec{r},t) = \sum_\lambda A(\lambda,t)\vec{w}_\lambda(\vec{r}); \qquad (2)$$

$$\int \vec{w}^*(\lambda',\vec{r})\vec{w}(\lambda,\vec{r})d\tau = \delta_{\lambda,\lambda'} \qquad (3)$$

and expanding the electrostatic potential $V^{(\alpha)}(\vec{r})$ of the electron-phonon interaction in terms of the partial potentials of the normal modes $v_\lambda(\vec{r})$:

$$V(\vec{r},t) = \sum_\lambda A(\lambda,t)v_\lambda(\vec{r}), \qquad (4)$$

we can represent the Maxwell equation in the form:

$$\mathrm{div}(\varepsilon_0 \varepsilon(\omega)\nabla v_\lambda(\vec{r})) = 0 \qquad (5)$$

with $\varepsilon(\omega) = \varepsilon(\infty)\dfrac{\omega^2 - \omega_{LO}^2}{\omega^2 - \omega_{TO}^2}$, and the Born-Huang equation in the form:

$$(\omega_\lambda^2 - \omega_{TO}^2)\vec{w}_\lambda = \omega_{TO}(\varepsilon_0(\varepsilon(0) - \varepsilon(\infty)))^{1/2}\nabla v_\lambda(\vec{r}), \qquad (6)$$

where $\varepsilon(\infty)$ and $\varepsilon(0)$ are the static and the high-frequency dielectric constants, respectively, $\omega_{TO}$ and $\omega_{LO}$ are the frequencies of the bulk transverse and longitudinal optical phonons. The parameters $\omega_{LO}$ and $\omega_{TO}$ of the dielectric function in the region of existence of the quasi-bulk modes remain constant; hence it follows from Eq. (5) that $\varepsilon_0 \varepsilon(\omega)\Delta v(q,\vec{r}) = 0$. For the longitudinal modes $\varepsilon(\omega = \omega_{LO}) = 0$ and $\Delta v \neq 0$. Under the condition that $v$ vanishes on the interfaces, we obtain from the Eq. (3) that $\Delta v = -q^2 v$. Using the orthonormalized set of solutions $\phi(q,\vec{r})$ of this eigenvalue problem, we can represent the sought functions in the form: $v(q,\vec{r}) = c(q)\phi(q,\vec{r})$. The coefficients $c(q)$ can be found using the condition (3) and Eq. (6). Substituting the quantized amplitudes $A(q)$ and the functions $v(q,\vec{r})$ in the expansion (4), we obtain the Hamiltonian of the interaction of the electron with quasi-bulk phonons:

$$H^b_{e-p} = \sum_q \gamma^{(b)}(q,\vec{r})(a_q + a_q^+), \qquad (7)$$

where

$$\gamma^{(b)}(q,\vec{r}) = -e\sqrt{\dfrac{\hbar\omega_{LO}}{2}\dfrac{1}{\varepsilon_0}\left(\dfrac{1}{\varepsilon(\infty)} - \dfrac{1}{\varepsilon(0)}\right)}\dfrac{\varphi(\vec{r},q)}{q}. \qquad (8)$$

It is worth mentioning that this formula is suitable in the isotropic case for any shape of the QD.

The interface modes are non-zero at the interfaces and the partial interface potentials are distributed in the whole space with boundary condition of vanishing at infinity. To obtain those potentials it is necessary to use the Maxwell equation in the general form (5) with the piecewise dielectric function $\varepsilon(\omega)$, which changes abruptly at the tetrahedral interfaces. The equation of the tetrahedron can be written in the form $ax + by + cz = \xi$, where $(a,b,c) = \pm 1$. The tetrahedral surfaces are defined as $\xi$=const (like the value r=const defines the surface of a sphere). Taking into account the tetrahedral symmetry of the shells, the coordinate dependence of the dielectric function can be written as $\varepsilon = \varepsilon(\xi)$.

Further, we represent the partial potential of the interface modes as the product:
$$v(q,\vec{r}) = \Phi(q,\xi)\psi(q,\vec{r}) \qquad (9)$$
in which the angular dependencies are described by the second factor. Substituting the definition (9) in the Maxwell equation and averaging it over the ensemble $\{q\}$ of orthonormal functions $\psi(q,\vec{r})$ at the tetrahedral surfaces $S(\xi)$, we find the equation for $\Phi(q,\xi)$. Solving this equation numerically by applying the condition of the existence of the non-zero solutions, the eigenfrequencies $\omega_j(q)$ can be found. For each $q$, there are five solutions. Accepting the spherical classification, we denote the frequencies of the interface modes as $\omega_j(l)$, where $l = s, p, d, ...$ and $j = 1,2,3,4,5$. The Hamiltonian of the interaction of the electron with interface phonons is represented in the form
$$H^{(I)}_{e\text{-}ph} = \sum_{j,q} \gamma^{(I)}_j(q)(a_{j,q} + a^\dagger_{j,q}), \qquad (10)$$
where
$$\gamma^{(I)}_j(l) = -e\left(\frac{\hbar}{2\omega_j(l)}\right)^{1/2} c_j(l)\Phi_j(l,\xi)\psi(l,\vec{r}), \quad (11)$$
$$c_j(l) = \left(\int \left(\frac{\omega_{TO}(\varepsilon_0(\varepsilon(0)-\varepsilon(\infty)))^{1/2}}{\omega_j^2(q)-\omega_{TO}^2}\right)^2 (\nabla \Phi_j \psi(l,\vec{r}))^2 d\tau\right)^{-1/2}$$
(12)

The Hamiltonian of the exciton-phonon interaction is then written as

$$H^{(\alpha)} = \sum_\alpha \gamma^{(\alpha)}_j(q,\vec{r}_e,\vec{r}_h)(a_j(q) + a^\dagger_j(q)),$$
$$\gamma^{(\alpha)}_{j,exc}(q,\vec{r}_e,\vec{r}_h) = \gamma^{(\alpha)}_j(q,\vec{r}_e) - \gamma^{(\alpha)}_j(q,\vec{r}_h) \quad [\alpha = b, I].$$
(13)

## 4. Spectrum of the photoluminescence of the CdS/HgS/CdS QDQW

Experiments [3] show that the PL spectrum of the QDQW corresponds to the equilibrium PL from the exciton ground state $\beta_0$. We use the formulas of Ref. [6] in the dipole approximation to calculate the intensities of the phonon lines in the PL spectrum of QDQW's.

In Fig. 2 the PL spectrum calculated for a single tetrahedral QDQW is compared with two experimental spectra, measured at different excitation wavelengths and intensities [7]. The experimentally determined Huang-Rhys parameter is 0.25±0.05 [3]. The dominant contribution to the intensity of the one-phonon lines in the calculated PL spectrum is due to p-like interface phonon modes. The two most intense one-phonon peaks correspond to the modes with amplitudes $V_{p,4}$ and $V_{p,5}$ and energies 34.3 meV and 36.7 meV, respectively. The average energy, weighted with the intensities of the two peaks, is 35.5 meV, what is in a good agreement with the experimentally determined value of 35.3±0.6 meV [3]. The insert to Fig. 2 shows the one- and two-phonon bands in the PL spectrum calculated within the adiabatic approximation. The intensities of the phonon peaks in the adiabatic approximation are dramatically lower than those calculated within the nonadiabatic theory [6].

In summary, we have shown a strong difference between the optical spectra of spherical and tetrahedral QDQW's. Only the simultaneous consideration of the tetrahedral shape of a QDQW, interface optical phonons, and nonadiabatic phonon-assisted transitions allows for profound understanding of the optical response of a QDQW. In particular, this analysis allowed us to interpret the observed PL spectra of a CdS/HgS/CdS QDQW.

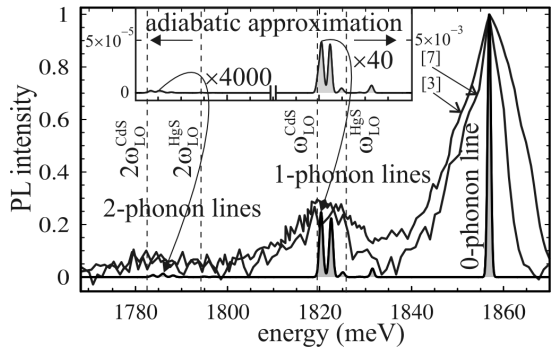

Fig.2. PL spectrum of a tetrahedral QDQW normalized to the zero-phonon line's intensity. Calculated spectral lines are broadened by a Gaussian to facilitate visualization. Experimental spectra are measured at T=10 K with excitation wavelength 633 nm and intensity 15 kW/cm² [3] and with excitation wavelength 442 nm and intensity 5 kW/cm² [7]. The insert gives one- and two-phonon bands (magnified by factors 40 and 4000, respectively) as calculated within the adiabatic approximation.

The authors thank T. Basché, who provided experimental data, F. Brosens, V. N. Gladilin, S. N. Klimin, A. A. Balandin and J. H. Wolter for fruitful discussions. This work was supported by the GOA BOF UA 2000, IUAP, FWO-V (Belgium), eiTT/COBRA at TU Eindhoven (the Netherlands), the MRDA-CRDF Award MP2-3044 (Moldova) and the EC GROWTH Programme, NANOMAT project G5RD-CT-2001-00545.